\documentstyle[12pt]{article}

\begin{document}
\begin{center}
{\Large{\bf Enhancement of Nuclear Spin Superradiance by Electron Resonator}
\footnote{Invited report at the International Workshop on Laser Physics,
Bordeaux (2000).} \\ [5mm]
V.I. Yukalov$^{*,**}$ and E.P. Yukalova$^{***}$} \\ [3mm]
{\it
$^{*}$Bogolubov Laboratory of Theoretical Physics \\
Joint Institute for Nuclear Research, Dubna 141980, Russia \\ [2mm]
$^{**}$International Institute of Theoretical and Applied Physics \\
Iowa State University, Ames, Iowa 50011-3022, USA \\ [2mm]
$^{***}$Laboratory of Informational Technologies \\
Joint Institute for Nuclear Research, Dubna 141980, Russia} \\ [20mm] 
\end{center}  

\begin{abstract}
Superradiance of nuclear spins is considered, when the nuclei interact via
hyperfine forces with electrons of a ferromagnet. The consideration is based
on a microscopic model. If the sample, coupled with a resonant electric
circuit, possesses electronic magnetization, then the electron subsystem
plays the role of an additional effective resonator, by enhancing the
coupling between nuclear spins and the resonant circuit. Radiation power can
be increased by three orders, while the radiation time of a superradiance
burst can become three times shorter. In the presence of dynamic nuclear
polarization, the regime of pulsing superradiance can occur.
\end{abstract}

\newpage

\section{Introduction}

The possibility of self-organized nuclear spin superradiance was predicted
by Bloembergen and Pound [1] and observed in a series of experiments [2--5].
A microscopic theory of this phenomenon was developed in Refs. [6--8], the
results being in good agreement with experiments as well as with computer
simulation [9--12].

The experiments [2--5] have dealt with proton spins inside paramagnets, such
as propanediol C$_3$H$_8$O$_2$, butanol C$_4$H$_9$OH, and ammonia NH$_3$.
When nuclei are incorporated in a ferromagnet, their spins interact with
those of electrons by means of hyperfine forces. If the electron subsystem
possesses long-range magnetic order, this can essentially influence nuclear
spin dynamics [13]. The aim of this report is to describe how the arising
electron ferromagnetism influences nuclear spin superradiance.

The description of nonlinear spin dynamics is based on the {\it scale
separation approach} [7,8,14]. Because of the importance of this approach
for obtaining a detailed picture of spin evolution, we find it pertinent to
briefly sketch in this introduction the main points of that technique as
applied to spin systems. The basic parts of the scale separation approach
[7,8,14] are: (i) Short-range stochastic quantization; (ii) Classification
of relative quasi-invariants; and (iii) Generalized averaging technique.

{\it Short-range stochastic quantization} makes it possible to derive a
closed set of evolution equations for a given statistical system. This is
achieved by decoupling correlators in such a way that takes into account
short-range incoherent effects. For illustration, let us consider spin
operators ${\bf S}_i$ labelled by a site index $i=1,2,\ldots,N$. The evolution
equations for the averages of these operators involve, as is known, binary
spin correlators. To render the system of equations closed, one could decouple
the binary correlators into the products of spin averages. This is what is
called the mean-field approximation or semiclassical approximation. Such
approximations take into account only long-range correlations but completely
ignore short-range effects. The latter, however, can be principally important
for the correct description of evolution. To take into account short-range
correlations, it is necessary to resort to a more elaborate decoupling of
spin correlators, than the mean-field one.

Let us present a binary spin correlator $<S_i^\alpha S_j^\beta>$, with
$i\neq j$, in the form
$$
<S_i^\alpha S_j^\beta>\; = \; < S_i^\alpha><S_j^\beta>\; + \;
<S_i^\alpha>\; \delta S_j^\beta + \; <S_j^\beta>\; \delta S_i^\alpha \; .
$$
Here the factors $<S_i^\alpha>$ are associated with long-range effects, in
many cases permitting the usage of the uniform approximation
$$
<S_i^\alpha> \; = \frac{1}{N}\; \sum_{i=1}^N \; <S_i^\alpha>\; .
$$
The terms $\delta S_i^\alpha$ describe short-range effects. For this purpose,
these terms are treated as random variables modelling local spin fluctuations.
To concretize the choice of the random variables, one has either to introduce
a related distribution or to define the first moments of the random-variable
products. If $\delta S_i^\alpha$ are assumed to be Gaussian variables, then
one has to define just the first two moments. Denoting the averaging over the
random variables $\delta S_i^\alpha$ as $\ll\ldots\gg$, we set
$$
\ll \delta S_i^\alpha\gg \; =0 \; .
$$
The binary correlators $\ll\delta S_i^\alpha\delta S_j^\beta\gg$ are to be
defined according to the nature of the stochastic variables $\delta S_i^\alpha$
describing local spin fluctuations [15]. For instance, we may set
$$
\ll \delta S_i^\alpha\delta S_j^\beta\gg \; =
\frac{1}{3}\; S(S+1)\delta_{ij}\delta_{\alpha\beta} \; ,
$$
where $S$ is a spin value.

The evolution equations for the spin components $S_i^\alpha$ are obtained
by means of the Heisenberg equations. Averaging these equations, with a
statistical operator $\hat\rho(0)$, one gets
$$
\frac{d}{dt}\; < S_i^\alpha> \; = -\; \frac{i}{\hbar}\; <\left [
S_i^\alpha,\; \hat H\right ]> \; ,
$$
where $\hat H$ is the system Hamiltonian. The right-hand side of this
equation contains binary spin correlators which are to be presented
according to the procedure of short-range stochastic quantization described
above. The resulting equation includes the stochastic terms
$\delta S_i^\alpha$, hence, this is a stochastic differential equation. In
this way, we obtain a closed system of equations although the price for this
is that the evolution equations become stochastic. The random variables
are associated with short-range, quantum, incoherent effects. That is why 
the procedure of deriving such equations can be called short-range 
stochastic quantization.

The notion of {\it relative quasi-invariants} [15] is introduced for solving
nonlinear systems of stochastic equations in partial derivatives. Let us
have a set of functions $f_n=f_n({\bf x},\xi,\varepsilon)$ depending on a
collection ${\bf x}=\{x_1,x_2,\ldots\}$ of variables $x_i$, on a set of
stochastic variables, $\xi$, and on an ensemble 
$\varepsilon=\{\varepsilon_1,\varepsilon_2,\ldots\}$ of small parameters, 
$|\varepsilon_k|\ll 1$. The collection ${\bf x}$ can contain, e.g.,
spatial and a temporal variables. For each variable $x_i$, we define a 
{\it variation length} $L_i$ as a characteristic distance at which the 
function $f_n$ changes essentially. If the function $f_n$ is periodic with 
respect to $x_i$, then the variation length $L_i$ is the distance between 
a maximum and a minimum of $f_n$ for varying $x_i$, that is, $L_i$ is a 
half-period. For a nonperiodic function, the variation length $L_i$ can be 
defined as the linear size of the considered system with respect to the 
variable $x_i$. In particular, $L_i$ can be infinite. Let us introduce the 
notation
$$
[f]_i \equiv \int\; \ll f({\bf x},\xi,\varepsilon)\gg \;
\prod_{j(\neq i)} \; \frac{dx_j}{L_j} \; ,
$$
in which the integration with respect to $x_j$ is over the interval $[0,L_j]$,
and $\ll\dots\gg$ implies a stochastic averaging over $\xi$. A function $f_m$
is an $x_i$-quasi-invariant with respect to $f_n$ if, and only if,
$$
\lim_{\varepsilon\rightarrow 0} \left [ \frac{\partial}{\partial x_i}\;
f_m\right ]_i = 0 \; , \qquad
\lim_{\varepsilon\rightarrow 0} \left [ \frac{\partial}{\partial x_i}\;
f_n\right ]_i \neq 0 \; ,
$$
where the limit $\varepsilon\rightarrow 0$ denotes that either all
$\varepsilon_k\rightarrow 0$ or that there exists a subset $\{\varepsilon_k\}$
of the set $\varepsilon$, such that the above limits hold true for all
$\varepsilon_k$ from this subset.

The notion of relative quasi-invariants makes it possible to classify the
solutions of evolution equations as fast or slow with respect to each other
and to generalize the averaging technique [16] to nonlinear systems of
stochastic equations in partial derivatives.

\section{Electron-Nuclear Hamiltonian}

We consider a Hamiltonian of the general form
\begin{equation}
\label{1}
\hat H = \hat H_e + \hat H_n + \hat H_{en} \; ,
\end{equation}
describing interacting electrons and nuclei. In the electron Hamiltonian
\begin{equation}
\label{2}
\hat H_e = -\; \frac{1}{2}\; \sum_{i\neq j} \; J_{ij}\; {\bf S}_i\cdot
{\bf S}_j - \mu_e\; \sum_i \; {\bf B}\cdot{\bf S}_i \; ,
\end{equation}
where $J_{ij}$ is an exchange interaction; ${\bf S}_i$, an electron spin
operator; $\mu_e=g_e\mu_B$, with $g_e$ being the electronic gyromagnetic
ratio and $\mu_B$, the Bohr magneton; ${\bf B}$ is a magnetic field. The
nuclear Hamiltonian
\begin{equation}
\label{3}
\hat H_n =\frac{1}{2}\; \sum_{i\neq j} \; \sum_{\alpha\beta} \;
C_{ij}^{\alpha\beta}\; I_i^\alpha\; I_j^\beta - \mu_n \;
\sum_i \; {\bf B}\cdot{\bf I}_i
\end{equation}
contains the dipole interactions
$$
C_{ij}^{\alpha\beta} =\frac{\mu_n^2}{r_{ij}^3}\; \left (
\delta_{\alpha\beta} - 3n_{ij}^\alpha\; n_{ij}^\beta\right ) \; ,
$$
where $\mu_n=g_n\mu_N$, with $g_n$ being the nuclear gyromagnetic ratio; 
$\mu_N$ is the nuclear magneton; $r_{ij}\equiv|{\bf r}_{ij}|$, 
${\bf r}_{ij}={\bf r}_i-{\bf r}_j$, ${\bf n}_{ij}\equiv{\bf r}_{ij}/r_{ij}$;  
and ${\bf I}_i$ is a nuclear spin operator. The electron-nuclear 
interactions are described by a hyperfine Hamiltonian
\begin{equation}
\label{4}
\hat H_{en} = A\; \sum_i\; {\bf S}_i\cdot{\bf I}_i + \frac{1}{2}\;
\sum_{i\neq j}\; \sum_{\alpha\beta}\; A_{ij}^{\alpha\beta}\; S_i^\alpha\;
I_j^\beta \; ,
\end{equation}
in which $A$ is an isotropic contact interaction and
$$
A_{ij}^{\alpha\beta} = \frac{\mu_e\mu_n}{r_{ij}^3}\; \left (
\delta_{\alpha\beta} - 3n_{ij}^\alpha\; n_{ij}^\beta \right )
$$
is a dipole interaction between electron and nuclear spins. The total field
\begin{equation}
\label{5}
{\bf B} = H_0{\bf e}_z + H_1{\bf e}_x \; , \qquad H_1 = H_a + H
\end{equation}
consists of an external field $H_0$ and the field $H_1$, in which $H_a$ is
an effective field of a transverse magnetocrystalline anisotropy and $H$ is a
resonator feedback field.

The sample is coupled to a resonant electric circuit of inductance $L$,
capacity $C$, and resistance $R$ defining the natural frequency $\omega$,
ringing time $\gamma_3$, and quality factor $Q$ by the relations
$$
\omega\equiv \frac{1}{\sqrt{LC}} \; , \qquad \gamma_3\equiv
\frac{\omega}{2Q}\; , \qquad Q\equiv \frac{\omega L}{R} \; .
$$
The electric current in the resonator coil is induced by the motion of the
transverse magnetization
\begin{equation}
\label{6}
M_x =\frac{1}{V}\; \sum_i\; \left ( \mu_e\; < S_i^x> \; +
\mu_n\; < I_i^x>\right ) \; ,
\end{equation}
where $V$ is the sample volume. The resonator feedback field is described by
the Kirchhoff equation
\begin{equation}
\label{7}
\frac{dH}{dt} +2\gamma_3 H +\omega^2  \int_0^t  H(\tau)\; d\tau +
4\pi\eta\; \frac{dM_x}{dt} = 0 \; ,
\end{equation}
where $\eta$ is a coil filling factor.

To write down the evolution equations, it is convenient to pass to the ladder
spin operators
$$
S_j^\pm \equiv S_j^x\pm iS_j^y \; , \qquad I_j^\pm \equiv I_j^x\pm iI_j^y\; .
$$
Then the electron spin Hamiltonian (2) takes the form
\begin{equation}
\label{8}
\hat H_e = -\; \frac{1}{2}\; \sum_{i\neq j} \; J_{ij}\left (
S_i^+\; S_j^- + S_i^z\; S_j^z\right ) -
\mu_e\; \sum_i\left [ H_0\; S_i^z +\frac{1}{2}\; H_1\left ( S_i^+ +
S_i^-\right )\right ] \; .
\end{equation}
The nuclear Hamiltonian (3) becomes
$$
\hat H_n =\frac{1}{2}\; \sum_{i\neq j}\; \left ( e_{ij}\; I_i^+\; I_j^- +
a_{ij}\; I_i^z\; I_j^z + b_{ij}\; I_i^+\; I_j^+ + b_{ij}^*\; I_i^-\; I_j^- +
2c_{ij}\; I_i^+\; I_j^z + \right.
$$
\begin{equation}
\label{9}
+ \left. 2c_{ij}^*\; I_i^-\; I_j^z\right ) -
\mu_n\; \sum_i\left [ H_0\; I_i^z + \frac{1}{2}\; H_1\left (
I_i^+ + I_i^-\right ) \right ] \; ,
\end{equation}
where
$$
a_{ij} \equiv C_{ij}^{zz}\; , \qquad e_{ij} \equiv \frac{1}{2}\left (
C_{ij}^{xx} + C_{ij}^{yy}\right ) \; ,
$$
$$
b_{ij}\equiv \frac{1}{4}\left ( C_{ij}^{xx} - C_{ij}^{yy} -
2i C_{ij}^{xy}\right ) \; , \qquad c_{ij} \equiv \frac{1}{2}\left (
C_{ij}^{xz} - iC_{ij}^{yz}\right ) \; .
$$
The hyperfine Hamiltonian (4) transforms to
$$
\hat H_{en} =\frac{1}{2}\; A\; \sum_i \left ( S_i^+\; I_i^- + S_i^-\; I_i^+
+2 S_i^z\; I_i^z\right ) +
$$
$$
+\frac{1}{2}\; \sum_{i\neq j} \left [ \frac{1}{2}\; e_{ij}'\left (
S_i^+\; I_j^- + S_i^-\; I_j^+\right ) + a_{ij}'\; S_i^z\; I_j^z +
b_{ij}'\; S_i^+\; I_j^+ + (b_{ij}')^*\; S_i^-\; I_j^- +\right.
$$
\begin{equation}
\label{10}
\left. + c_{ij}'\left ( S_i^+\; I_j^z + S_i^z\; I_j^+\right ) +
(c_{ij}')^*\left ( S_i^-\; I_j^z + S_i^z\; I_j^-\right )\right ]\; ,
\end{equation}
with
$$
a_{ij}'\equiv A_{ij}^{zz} \; , \qquad e_{ij}'\equiv \frac{1}{2}\left (
A_{ij}^{xx}+A_{ij}^{yy}\right ) \; ,
$$
$$
b_{ij}'\equiv \frac{1}{4}\left ( A_{ij}^{xx}-A_{ij}^{yy} -2iA_{ij}^{xy}
\right ) \; , \qquad c_{ij}'\equiv \frac{1}{2}\left ( A_{ij}^{xz} -
iA_{ij}^{yz} \right ) \; .
$$

\section{Evolution Equations}

From the Heisenberg equations for the spin operators we derive the
equations of motion for the ladder and longitudinal spins. For the ladder
electron spins, we have
$$
i\; \frac{dS_i^-}{dt} = -\mu_e\left ( H_0\; S_i^- - H_1\; S_i^z\right ) -
\sum_{j(\neq i)}\; J_{ij}\left ( S_i^-\; S_j^z - S_i^z\; S_j^-\right ) +
$$
$$
+ A(S_i^-\; I_i^z - S_i^z\; I_i^-) +
\frac{1}{2}\; \sum_{j(\neq i)}\left [ a_{ij}'\; S_i^-\; I_j^z -
e_{ij}'\; S_i^z\; I_j^- - 2b_{ij}'\; S_i^z\; I_j^+ \right. +
$$
\begin{equation}
\label{11}
+ \left. c_{ij}'\; S_i^-\; I_j^+ + (c_{ij}')^*\; S_i^-\; I_j^- -
2c_{ij}'\; S_i^z\; I_j^z\right ] \; ,
\end{equation}
where and in what follows we set, for the simplicity of notations, the Plank
constant $\hbar\equiv 1$. An equation for $S_i^+$ is obtained from Eq. (11)
by means of Hermitian conjugation. For the longitudinal electron spins, we get
$$
i\; \frac{dS_i^z}{dt} =\frac{1}{2}\;\mu_e\; H_1 \left ( S_i^- - S_i^+
\right ) +
$$
$$
+ \frac{1}{2}\; \sum_{j(\neq i)} \; J_{ij}\left ( S_i^-\; S_j^+ -
S_i^+\; S_j^-\right ) +
\frac{1}{2}\; A\left ( S_i^+\; I_i^- - S_i^-\; I_i^+\right ) +
$$
$$
+\frac{1}{2}\; \sum_{j(\neq i)}\left [ \frac{1}{2}\; e_{ij}'\left (
S_i^+\; I_j^- - S_i^-\; I_j^+\right )  + b_{ij}'\; S_i^+\; I_j^+ \right. -
$$
\begin{equation}
\label{12}
- \left. (b_{ij}')^*\; S_i^-\; I_j^- + c_{ij}'\; S_i^+\; I_j^z -
(c_{ij}')^*\; S_i^-\; I_j^z\right ] \; .
\end{equation}
For the ladder nuclear spins, we find
$$
i\; \frac{dI_i^-}{dt} = -\mu_n\left ( H_0\; I_i^- - H_1\; I_i^z\right ) +
$$
$$
+ \sum_{j(\neq i)}\; \left (a_{ij}\; I_i^-\; I_j^z - e_{ij}\; I_i^z\; I_j^-
-2b_{ij}\; I_i^z\; I_j^+ + c_{ij}\; I_i^-\; I_j^+ + c_{ij}^*\;
I_i^-\; I_j^- - 2c_{ij}\; I_i^z\; I_j^z \right ) +
$$
$$
+ A(I_i^-\; S_i^z - I_i^z\; S_i^-)
+\frac{1}{2}\; \sum_{j(\neq i)}\left [ a_{ij}'\; I_i^-\; S_j^z -
e_{ij}'\; I_i^z\; S_j^- - 2b_{ij}'\; I_i^z\; S_j^+ \right. +
$$
\begin{equation}
\label{13}
+\left. c_{ij}'\; I_i^-\; S_j^+ + (c_{ij}')^*\; I_i^-\; S_j^- -
2c_{ij}'\; I_i^z\; S_j^z\right ] \; .
\end{equation}
Finally, for the longitudinal nuclear spins, we obtain
$$
i\; \frac{dI_i^z}{dt} = \frac{1}{2}\; \mu_n\;H_1\left( I_i^- - I_i^+\right ) +
$$
$$
+\sum_{j(\neq i)}\; \left [ \frac{1}{2}\; e_{ij}\left ( I_i^+\; I_j^-
 - I_i^-\; I_j^+\right ) + b_{ij}\; I_i^+\; I_j^+ - b_{ij}^*\;
I_i^-\; I_j^- + c_{ij}\; I_i^+\; I_j^z - c_{ij}^*\; I_i^-\; I_j^z \right ]
+
$$
$$
+\frac{1}{2}\; A\left ( I_i^+\; S_i^-  - I_i^-\; S_i^+ \right )
+ \frac{1}{2}\; \sum_{j(\neq i)} \left [\frac{1}{2}\; e_{ij}'\;
\left ( I_i^+\; S_j^- - I_i^-\; S_j^+\right ) + b_{ij}'\; I_i^+\; S_j^+\right.-
$$
\begin{equation}
\label{14}
-\left. (b_{ij}')^*\; I_i^-\; S_j^- + c_{ij}'\; I_i^+\; S_j^z -
(c_{ij}')^*\; I_i^-\; S_j^z \right ] \; .
\end{equation}

In the calculations below, we shall employ the following properties of
the dipole interactions:
$$
\sum_\alpha \; A_{ij}^{\alpha\alpha} = \sum_\alpha \;
C_{ij}^{\alpha\alpha} = 0\; , \qquad
\sum_{j(\neq i)}\; A_{ij}^{\alpha\beta} = \sum_{j(\neq i)}\;
C_{ij}^{\alpha\beta} = 0 \; .
$$
From the first of these properties, it follows that
$$
e_{ij}=-\; \frac{1}{2}\; a_{ij} \; , \qquad e_{ij}' = -\; \frac{1}{2}\;
a_{ij}' \; .
$$
The second of the summation properties above is, strictly speaking,
approximate being valid up to boundary effects.

The Zeeman frequencies for electrons and nuclei are
\begin{equation}
\label{15}
\omega_e \equiv \mu_e\; H_0 \; , \qquad \omega_n\equiv \mu_n\; H_0\; ,
\end{equation}
respectively. The characteristic wavelengths $2\pi/\omega_e$ and
$2\pi/\omega_n$ are much larger than the mean distance between spins.
Therefore, for the statistical averages of spin operators of electrons,
\begin{equation}
\label{16}
x\equiv \; < S_i^-> \; , \qquad z\equiv \; <S_i^z> \; ,
\end{equation}
and nuclei,
\begin{equation}
\label{17}
u\equiv \; <I_j^->\; , \qquad s\equiv\; <I_j^z>\; ,
\end{equation}
we may use the uniform approximation. At the same time, local spin
fluctuations, disturbing space uniformity, will be taken into account by
means of the short-range stochastic quantization explained in the
Introduction. Realizing this procedure, we come to the following expressions
of local random fields:
$$
\xi_0 \equiv \frac{1}{2}\; \sum_{j(\neq i)} \left [ 2J_{ij}\left (
\delta S_i^z -\delta S_j^z\right ) + a_{ij}'\delta I_j^z +
c_{ij}'\delta I_j^+ +(c_{ij}')^*\; \delta I_j^-\right ] \; ,
$$
$$
\xi\equiv \frac{1}{2}\; \sum_{j(\neq i)} \left [ 2J_{ij}\left (
\delta S_i^- - \delta S_j^-\right ) + e_{ij}'\;\delta I_j^- +
2b_{ij}'\;\delta I_j^+ + 2c_{ij}'\;\delta I_j^z\right ] \; ,
$$
$$
\varphi_0 \equiv \frac{1}{2}\; \sum_{j(\neq i)} \; \left [
3a_{ij}\;\delta I_j^z + 2c_{ij}\;\delta I_j^+ + 2c_{ij}^*\;\delta I_j^- +
a_{ij}'\; \delta S_j^z + c_{ij}'\; \delta S_j^+ + (c_{ij}')^*\;
\delta S_j^-\right ] \; ,
$$
\begin{equation}
\label{18}
\varphi \equiv \frac{1}{2}\; \sum_{j(\neq i)} \; \left [ 4b_{ij}\;
\delta I_j^+ + 4c_{ij}\;\delta I_j^z + e_{ij}'\;\delta S_j^- +
2b_{ij}'\;\delta S_j^+ + 2c_{ij}'\;\delta S_j^z\right ] \; .
\end{equation}
Introduce also the anisotropy frequencies
\begin{equation}
\label{19}
\alpha_e \equiv \mu_e\; H_a \; , \qquad \alpha_n\equiv \mu_n\; H_a \; .
\end{equation}
Averaging the Heisenberg equations (11) to (14), we employ the short-range
stochastic quantization, include into the equations the transverse and  
longitudinal spin relaxation parameters, and envolve the notations (15) to
(19). This results in the evolution equations for the electron transverse
spin average,
\begin{equation}
\label{20}
\frac{dx}{dt} = i (\omega_e - A\; s -\xi_0 +i\gamma_2) x -
i(\alpha_e - A\; u -\xi + \mu_e\; H) z \; ,
\end{equation}
electron longitudinal spin average,
\begin{equation}
\label{21}
\frac{dz}{dt} =\frac{i}{2}\; \left ( \alpha_e - A\; u -\xi +\mu_e\; H
\right ) x^* -\; \frac{1}{2} \left ( \alpha_e - A\; u^* -\xi^*
+\mu_e\; H \right ) x -\gamma_1(z -\sigma) \; ,
\end{equation}
where $\sigma$ is a stationary electron single-spin polarization, for the
electron transverse modulus of spin squared,
\begin{equation}
\label{22}
\frac{d |x|^2}{dt} = -2\gamma_2|x|^2 + i(\alpha_e - A\; u^* -\xi^* +
\mu_e\; H) z x - i(\alpha_e - A\; u -\xi +\mu_e\; H) z x^* \; ,
\end{equation}
for the transverse nuclear spin average,
\begin{equation}
\label{23}
\frac{du}{dt} = i(\omega_n - A\; z -\varphi_0 +i\Gamma_2) u -
i(\alpha_n - A\; x-\varphi +\mu_n\; H) s \; ,
\end{equation}
longitudinal nuclear spin average,
\begin{equation}
\label{24}
\frac{ds}{dt} =\frac{i}{2}\left ( \alpha_n - A\; x -\varphi +\mu_n\; H
\right ) u^* -\; \frac{i}{2}\left ( \alpha_n - A\; x^* -\varphi^*
+\mu_n\; H\right ) u - \Gamma_1 (s-\zeta) \; ,
\end{equation}
where $\zeta$ is a stationary nuclear spin polarization, and, finally, for
the nuclear transverse modulus of spin squared,
\begin{equation}
\label{25}
\frac{d|u|^2}{dt} = -2\Gamma_2\; |u|^2 + i \left ( \alpha_n - A\; x^* -
\varphi^* + \mu_n\; H\right ) s u - i\left ( \alpha_n - A\; x -
\varphi + \mu_n\; H\right )  s u^* \; .
\end{equation}
The evolution equations (20) to (25) form a set of nonlinear stochastic
differential equations, which are also to be complemented by the Kirchhoff
equation (7), where the magnetization (6) has to be understood as the total
average
\begin{equation}
\label{26}
M_x =\frac{1}{2}\; \ll \left [ \mu_e\; \rho_e (x + x^*) + \mu_n\; \rho_n
(u + u^*)\right ] \gg \; ,
\end{equation}
with $\rho_e$ and $\rho_n$ being the electron and nuclear densities,
respectively. It is convenient to rewrite the feedback equation (7) in the
form
\begin{equation}
\label{27}
H= -4\pi\; \eta\; \int_0^t \; G(t-\tau)\; dM_x(\tau) \; ,
\end{equation}
with the Green function
$$
G(t) =\left ( \cos\omega_3 t -\; \frac{\gamma_3}{\omega_3}\;
\sin\omega_3 t\right )\; e^{-\gamma_3 t} \; , \qquad
\omega_3\equiv \sqrt{\omega^2 -\gamma_3^2} \; .
$$

To make the set of the evolution equations completely defined, we need to
concretize the random variables (18). For this purpose, we treat these
variables as Gaussian, with the stochastic averages
\begin{equation}
\label{28}
\ll \xi_0\gg \; = \; \ll \xi\gg\; = 0 \; , \qquad
\ll \xi_0\;\xi\gg \; = 0 \; , \qquad
\ll \xi_0^2\gg \; =\; \ll |\xi|^2\gg \; =\gamma_*^2
\end{equation}
for the electron-spin local fluctuations, and with
\begin{equation}
\label{29}
\ll \varphi_0\gg \; = \; \ll \varphi\gg\; = 0 \; , \qquad
\ll \varphi_0\;\varphi\gg \; = 0 \; , \qquad
\ll \varphi_0^2\gg \; =\; \ll |\varphi|^2\gg \; =\Gamma_*^2
\end{equation}
for the nuclear-spin local fluctuations. The quantities $\gamma_*$ and
$\Gamma_*$ are
inhomogeneous widths, for which, according to the definition (18), one may
write
\begin{equation}
\label{30}
\gamma_*^2 = \Gamma_{ee}^2 + \Gamma_{en}^2 \; , \qquad
\Gamma_*^2 = \Gamma_{en}^2 + \Gamma_{nn}^2 \; ,
\end{equation}
where the corresponding terms are due to electron-electron, electron-nuclear,
and to nuclei local interactions.

\section{Small Parameters}

To simplify the evolution equations (20) to (25), let us take into account
the existence of several small parameters. First, the longitudinal and
transverse relaxation parameters are assumed to be small as compared to the
Zeeman frequencies, so that
\begin{equation}
\label{31}
\frac{\gamma_1}{\omega_e} \ll 1 \; , \qquad
\frac{\gamma_2}{\omega_e} \ll 1 \; , \qquad
\frac{\Gamma_1}{\omega_n} \ll 1 \; , \qquad
\frac{\Gamma_2}{\omega_n} \ll 1 \; .
\end{equation}
Inhomogeneous broadening is supposed also to be week,
\begin{equation}
\label{32}
\frac{\gamma_*}{\omega_e} \ll 1 \; , \qquad
\frac{\Gamma_*}{\omega_n} \ll 1 \; .
\end{equation}
The same concerns the magnetocrystalline anisotropy,
\begin{equation}
\label{33}
\frac{\alpha_e}{\omega_e} \ll 1 \; , \qquad
\frac{\alpha_n}{\omega_n} \ll 1 \; .
\end{equation}
The energy of the spin interaction with the resonator feedback field is 
much weeker than the corresponding Zeeman frequencies, which implies that 
$$
\frac{\rho_e\;\mu_e^2}{\omega_e} \ll 1 \; , \qquad
\frac{\rho_n\;\mu_e\;\mu_n}{\omega_e} \ll 1 \; ,
$$
\begin{equation}
\label{34}
\frac{\rho_e\;\mu_e\;\mu_n}{\omega_n}\ll 1 \; , \qquad
\frac{\rho_n\;\mu_n^2}{\omega_n} \ll 1 \; .
\end{equation}
This is equivalent to saying that $|\mu_e\; H|\ll\omega_e$ and
$|\mu_n\; H|\ll\omega_n$. The contact hyperfine interaction is smaller than
the electron Zeeman frequency,
\begin{equation}
\label{35}
\frac{A}{\omega_e} \ll 1 \; ,
\end{equation}
but $A$ can be comparable or even much larger than $\omega_n$. The resonator
is of good quality, i.e. its quality factor is high, $Q\gg 1$, which, because
of the relation $\gamma_3\equiv\omega/2Q$, means that the resonator ringing
time is small,
\begin{equation}
\label{36}
\frac{\gamma_3}{\omega} \ll 1 \; .
\end{equation}
Since the nuclear magneton is three orders smaller than the Bohr magneton, one
has
\begin{equation}
\label{37}
\frac{\mu_n}{\mu_e} \ll 1 \; .
\end{equation}
And also, one usually has
\begin{equation}
\label{38}
\frac{\Gamma_1}{\gamma_1} \ll 1 \; , \qquad
\frac{\Gamma_2}{\gamma_2} \ll 1 \; .
\end{equation}
Due to the inequality (37), the Zeeman frequencies are related as
\begin{equation}
\label{39}
\frac{\omega_n}{\omega_e} \ll 1 \; .
\end{equation}
Therefore, the resonant electric circuit can be tuned either to $\omega_n$
or $\omega_e$.

In general, the dynamics of electron spins is similar to that of nuclear
spins. The main difference is that nuclear spins, owing to the inequality
(37), or to that (39), weekly influence the behaviour of electron spins.
While, to the contrary, electronic spins, can essentially influence spin
dynamics. For instance, the effective electronic spin-resonance frequency,
as is seen from Eq. (20), is shifted as $\omega_e-As$, as a result of the
hyperfine interaction. But due to the inequality (35), this shift is very
small, and can be neglected. To the contrary, the effective nuclear
magnetic-resonance frequency, as follows from Eq. (23), is shifted as
$\omega_n-Az$, which is a kind of the dynamical frequency shift [17,18].
In the presence of the long-range magnetic order in the electron subsystem,
the nuclear magnetic-resonance frequency becomes
\begin{equation}
\label{40}
\omega_N \equiv \omega_n - A\;\sigma \; ,
\end{equation}
where $\sigma$ is a stationary electron magnetization. Since electrons,
especially those possessing magnetic order, can essentially influence the
evolution of nuclear spins, but not conversely, the nuclear spin dynamics
exhibits more varieties and is more interesting than that of electronic
spins. Therefore, we shall concentrate on the nuclear spin dynamics, and will
imply in what follows that the resonator is tuned to the magnetic resonance
frequency (40), so that the quasiresonance condition
\begin{equation}
\label{41}
\left | \frac{\Delta_N}{\omega_N}\right | \ll 1 \; , \qquad
\Delta_N \equiv \omega -\omega_N
\end{equation}
holds true.

Experimentally, the radiation intensity of moving spins can be observed
through measuring the current power
$$
P \equiv R\; J^2 \; , \qquad J^2 =\frac{V\;H^2}{4\pi\;\eta\; L} \; ,
$$
where $R$ and $L$ are resistivity and inductance of the resonant electric
circuit. Thus, we have
\begin{equation}
\label{42}
P =\frac{\gamma_3\; V}{2\pi\;\eta}\; H^2 \; .
\end{equation}
One may analyze how the arising magnetic order changes the power (42) or
its time average $\overline P$, with an averaging over fast oscillations.
Considering $P=P(\sigma)$ as a function of electron magnetization, one can
study the relative difference
\begin{equation}
\label{43}
\delta P(\sigma) \equiv \frac{P(\sigma)-P(0)}{P(0)} \; .
\end{equation}
In order to estimate the values (42) and (43), one has to calculate the
resonator field (27), which, under the inequality (36), envolves the Green
function
$$
G(t) = \cos(\omega\; t)\; e^{-\gamma_3\; t} \; .
$$
Estimates show [13] that the relative quantity (43) is
\begin{equation}
\label{44}
\delta P(\sigma) \sim \;
\frac{\rho_e\;\mu_e\; A\; \sigma}{\rho_n\;\mu_n\;\omega_N} \; .
\end{equation}
Hence, if $\rho_e\sim\rho_n$ and $\omega_N\sim A$, the power (42) can be
increased by three orders when $\sigma\sim 1$, since then
\begin{equation}
\label{45}
\delta P(\sigma) \sim \; \frac{\mu_e}{\mu_n}\; \sigma \gg 1 \; ,
\end{equation}
which follows from inequality (37). This is the {\it enhancement effect}
caused by the electron magnetization.

\section{Time Evolution}

The existence of small parameters, discussed in the previous section, allows
us to classify the solutions to the evolution equations (20) to (35) onto
fast or slow as compared to each other. Thus, among all functions, $x$
is the fastest one. In other words, all solutions $z,\; |x|^2,\; u,\; s$,
and $|u|^2$ are temporal quasi-invariants with respect to $x$. Dynamics of
nuclear spins is described by equations (23) to (25). Averaging these
equations over the period $T_e\equiv 2\pi/\omega_e$ of the fastest
oscillations, related to $x$, and setting
$$
\frac{1}{T_e}\; \int_t^{t+T_e}  x(\tau)\; d\tau =0 \; , \qquad
\frac{1}{T_e}\; \int_t^{t+T_e}  z(\tau)\; d\tau =\sigma \; ,
$$
we have
$$
\frac{du}{dt} = i(\omega_N -\varphi_0 +i\Gamma_2 ) \; u -
i(\alpha_n -\varphi +\mu_n\; \overline H)\; s \; ,
$$
$$
\frac{ds}{dt} =\frac{i}{2}(\alpha_n -\varphi + \mu_n\; \overline H)\; u^* -
\frac{i}{2}(\alpha_n -\varphi^* +\mu_n\;\overline H)\; u -
\Gamma_1(s-\zeta)\; ,
$$
and similarly for $|u|^2$, where
$$
\overline H \equiv \frac{1}{T_e} \; \int_t^{t+T_e} \; H(\tau)\; d\tau \; .
$$
From here, the functions $s$ and $|u|^2$ are quasi-invariants with respect
to $u$.

Following further the averaging technique, one can derive the equations for
the slowest functions $s$ and $|u|^2$. The latter equations allow analytical
solutions in the case of short time $t\ll T_2^*=\Gamma_*^{-1}$. This
analysis can be found in Refs. [6--8,13]. The appearance of pure spin
superradiance depends on the value of an affective coupling between the
nuclear spins and the resonator and also on the level of the initial
spin polarization. The latter can be varied in a large diapason ranging
from zero to practically $100\%$ [19].

For larger times, when $t\sim T_2^*$ or $t\gg T_2^*$, it is necessary to
accurately consider the inhomogeneous broadening. Then the analytical
solutions of the evolution equations is not available and one has to resort
to numerical calculations. When the polarization of nuclear spins is
supported by the procedure of dynamical nuclear polarization, so that the 
parameter $\zeta <0$ in Eq. (24), then the regime of {\it pulsing spin
superradiance} can arise [20].

To illustrate the general behaviour of nuclear spin dynamics, we present
the results of numerical calculations. Figure 1 shows the regime of weak
superradiance [8] without dynamical nuclear polarization, while Figs. 2 to
8 demonstrate the regime of pulsing spin superradiance occurring in the
presence of dynamical polarization. Time is everywhere measured in units
of $\Gamma_2^{-1}$. In Figs. 2 to 8, the effective coupling parameter is set
$g=10$, and the pumping parameter is $\zeta=-0.5$. The function
$w(t)=|u(t)|^2$ is proportional to the current power. This is why we
concentrate our attention on this behaviour. We study the behaviour of $w(t)$
for the varying parameter $\gamma\equiv\Gamma_1/\Gamma_2$ and for different
initial conditions $w_0\equiv w(0)$ and $s_0\equiv s(0)$. To simplify
calculations, the resonator feedback field was treated in an approximation
that neglects inhomogeneous broadening, because of which the presented
figures should be considered as only a qualitative illustration. But we do
hope that the regime of pulsing superradiance will survive in a more elaborate
treatment that is in progress.

\vskip 5mm
{\parindent=0pt
{\bf Acknowledgement}}

\vskip 2mm

We appreciate support from the Bogolubov-Infeld Grant of the State Agency for
Atomic Energy, Poland, and from the Grant of the International Institute
of Theoretical and Applied Physics, Iowa State University, USA.

\newpage

\begin{center}

{\bf Figure Captions}

\end{center}

{\bf Fig. 1}. The dipole radiation intensity $I(t)$ of nuclear spins, the
coherence coefficient $C_{coh}(t)$, and the polarization $p_z(t)=-s(t)$ as
functions of time. The radiation intensity is given in arbitrary units, and
time is measured in units of $\Gamma_2^{-1}$. This figure shows the regime
of weak superradiance [8].

\vskip 5mm

{\bf Fig. 2}. The regime of pulsing spin superradiance for $\gamma=0.01$,
$w_0=10^{-6}$, and $s_0=-0.1$: (a) $w(t)$; (b) $s(t)$.

\vskip 5mm

{\bf Fig. 3}. The function $w(t)$ versus time for $\gamma=0.001$,
$w_0=10^{-6}$, and $s_0=-0.1$.

\vskip 5mm

{\bf Fig. 4}. The solution $w(t)$ as a function of time under the initial
conditions $w_0=0.1$ and $s_0=-0.25$ for varying $\gamma$: $\gamma=1$
(solid line) and $\gamma=0.5$ (dashed line).

\vskip 5mm

{\bf Fig. 5}. The function $w(t)$ at $w_0=0.5$ and $s_0=0.5$ for $\gamma=1$
(solid line) and $\gamma=0.5$ (dashed line).

\vskip 5mm

{\bf Fig. 6}. Pulsing spin superadiance for $\gamma=0.01$, $w_0=0.01$, and
$s_0=-0.1$.

\vskip 5mm

{\bf Fig. 7}. Pulsing superadiance in the case of the parameters
$\gamma=0.01$, $w_0=0.001$, and $s_0=-0.5$.

\vskip 5mm

{\bf Fig. 8}. The regime of pulsing spin superradiance for $\gamma=0.01$,
$w_0=0.1$, and $s_0=-0.1$.


\newpage

\begin{thebibliography}{99}

\bibitem{1}
Bloembergen, N. and Pound, R.V., 1954, {\it Phys. Rev.}, {\bf 95}, 8.

\bibitem{2}
Kiselev, J.F., Prudkoglyad, A.F., Shumovsky, A.S., and Yukalov, V.I.,
1988, {\it Mod. Phys. Lett. B}, {\bf 1}, 409.

\bibitem{3}
Kiselev, Y.F., Prudkoglyad, A.F., Shumovsky, A.S., and Yukalov, V.I.,
1988, {\it J. Exp. Theor. Phys.}, {\bf 67}, 413.

\bibitem{4}
Bazhanov, N.A. {\it et al}., 1990, {\it J. Exp. Theor. Phys.}, {\bf 70},
1128.

\bibitem{5}
Reichertz, L. {\it et al.}, 1994, {\it Nucl. Instrum. Methods Phys. Res. A},
{\bf 340}, 278.

\bibitem{6}
Yukalov, V.I., 1995, {\it Phys. Rev. Lett.}, {\bf 75}, 3000.

\bibitem{7}
Yukalov, V.I., 1995, {\it Laser Phys.}, {\bf 5}, 970.

\bibitem{8}
Yukalov, V.I., 1996, {\it Phys. Rev. B}, {\bf 53}, 9232.

\bibitem{9}
Belozerova, T.S., Henner, V.K., and Yukalov, V.I., 1992, {\it Phys. Rev. B},
{\bf 46}, 682.

\bibitem{10}
Belozerova, T.S., Henner, V.K., and Yukalov, V.I., 1992, {\it Laser Phys.},
{\bf 2}, 545.

\bibitem{11}
Belozerova, T.S., Davis, C.L., and Henner, V.K., 1998,
{\it Phys. Rev. B}, {\bf 58}, 3111.

\bibitem{12}
Belozerova, T.S., Davis, C.L., and Henner, V.K., 1999,
{\it Comput. Phys. Commun.}, {\bf 121}, 214.

\bibitem{13}
Yukalov, V.I., Cottam, M.G., and Singh, M.R., 1999, {\it Phys. Rev. B},
{\bf 60}, 1227.

\bibitem{14}
Yukalov, V.I., 1993, {\it Laser Phys.}, {\bf 3}, 870.

\bibitem{15}
Yukalov, V.I., 2000, {\it Proc. Int. Soc. Opt. Eng.}, {\bf 4061}, 2.

\bibitem{16}
Bogolubov, N.N. and Mitropolsky, Y.A., 1961, {\it Asymptotic Methods in the
Theory of Nonlinear Oscillations} (New York: Gordon and Breach).

\bibitem{17}
Turov, E.A. and Petrov, M.P., 1972, {\it Nuclear Magnetic Resonance in
Ferro-- and Antiferromagnets} (Jerusalem: Halsted).

\bibitem{18}
Khutsishvili, K.O. and Chkhaidze, S.G., 1992, {\it Physica B}, {\bf 176}, 54.

\bibitem{19}
Kisselev, Y.F., 2000, {\it Phys. Part. Nucl.}, {\bf 31}, 714.

\bibitem{20}
Yukalov, V.I. and Yukalova, E.P., 1998, {\it Laser Phys.}, {\bf 8}, 1029.

\end{thebibliography}
\end{document}